\documentclass[letterpaper, 10 pt, conference]{ieeeconf}  

\IEEEoverridecommandlockouts                              

\usepackage{graphicx} 
\usepackage{amsmath} 
\usepackage{amssymb}  
\usepackage{subfig}
\let\OLDthebibliography\thebibliography
\renewcommand\thebibliography[1]{
  \OLDthebibliography{#1}
  \setlength{\parskip}{0pt}
  \setlength{\itemsep}{0pt plus 0.3ex}
}

\newenvironment{itemize1}{
\begin{itemize}
  \setlength{\itemsep}{1pt}
  \setlength{\parskip}{0pt}
  \setlength{\parsep}{0pt}
  \setlength{\leftmargin}{0pt}
}{\end{itemize}}

\usepackage{fancyhdr} 
\title{\LARGE \bf
A co-design approach for a rehabilitation robot coach for physical rehabilitation based on the error classification of motion errors
}

\author{Maxime Devanne$^{1}$, Sao Mai Nguyen$^{1}$,Olivier Remy-Neris$^2$,  Beatrice Le Gales$^3$, Gilles Kermarrec$^3$, Andre Thepaut $^1$
\thanks{\footnotesize $^{1}$ IMT Atlantique, Lab-STICC, UBL, France
        {\tt \footnotesize maxime.devanne@imt-atlantique.fr, nguyensmai@gmail.com}}%
\thanks{\footnotesize $^{2}$CHRU de Brest, Hopital Morvan, Service de Medecine Physique et Readaptation, France}
\thanks{\footnotesize $^3$ European Research Center for Virtual Reality and Research Center for Education, 
Learning and Didactics, European University of Brittany, France }
\thanks{\footnotesize *The research work presented in this paper is partially supported by the EU FP7 grant ECHORD++ KERAAL and by 
the European Regional Fund (FEDER) via the VITAAL Contrat Plan Etat Region~\cite{thepaut:hal-01514898}.}
}

\begin{document}

\maketitle

\thispagestyle{empty}
\thispagestyle{fancy}
\lhead{}
\chead{\vspace{-40pt}
\texttt{\scriptsize{M. Devanne, S.M. Nguyen, ,O Remy-Neris, B Le Gales, G. Kermarrec and A. Thepaut , A co-design approach for a rehabilitation robot coach for physical rehabilitation based on the error classification of motion errors. IEEE International Conference on Robotic Computing (IRC) 2018. }}
\vspace{5pt}}
\rhead{}
\cfoot{}

\begin{abstract}
The rising number of the elderly incurs growing concern about healthcare, and in particular rehabilitation healthcare. Assistive technology and assistive robotics in particular may help to improve this process. We develop a robot coach capable of demonstrating rehabilitation exercises to patients, watch a patient carry out the exercises and give him feedback so as to improve his performance and encourage him. The HRI of the system is based on our study with a team of rehabilitation therapists and with the target population. 

The system relies on human motion analysis. We develop a method for learning a probabilistic representation of ideal movements from expert demonstrations. A Gaussian Mixture Model is employed from position and orientation features captured using a Microsoft Kinect v2. For assessing patients' movements, we propose a real-time multi-level analysis to both temporally and spatially identify and explain body part errors. This analysis combined with a classification algorithm allows the robot to provide coaching advice to make the patient improve his movements. The evaluation on three rehabilitation exercises shows the potential of the proposed approach for learning and assessing kinaesthetic movements.\end{abstract}

\section{INTRODUCTION}

Low back pain is a leading cause disabling people particularly affecting the elderly, whose proportion in European societies keeps rising, incurring growing concern about healthcare. 50 to 80\% of the world population suffers at a given moment from back pain which makes it in the lead in terms of health problems occurrence frequency~\cite{who2003burden}. 
To tackle this chronic low back pain, regular physical rehabilitation exercises is considered most effective~\cite{Kent2012T}.
%
%

\begin{figure}[!th]
\vspace{-0.2cm}
\centering
\includegraphics[width=0.8\columnwidth]{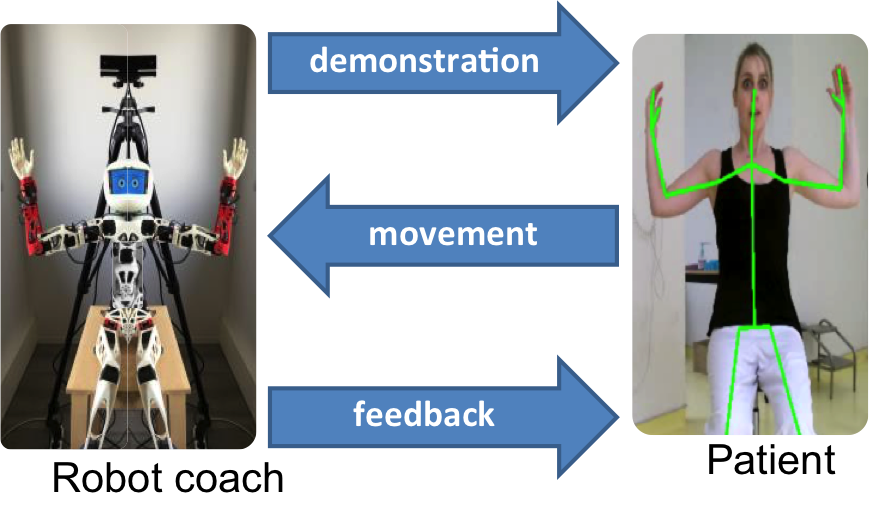}
\caption{\small Setting of the system including a Microsoft Kinect v2 and an open source humanoid robot called Poppy. After demonstrating the exercise, the robot can track the skeleton to analyse the movement in order to give feedback.}\label{overviewSystem}
\vspace{-0.2cm}
\end{figure}

With this perspective, solutions are being developed based on assistive technology and particularly robotics. In KERAAL project, we are developing a robot coach for physical rehabilitation exercises. The goal is to increase the time patients spend exercising, by alleviating the lack of time a physiotherapist can spend monitoring a patient \cite{Nguyen2016R2IISHRIC}. The system is composed of a low cost stereo vision camera (Microsoft Kinect v2) and the humanoid robot Poppy (Fig.~\ref{overviewSystem}). The Poppy robot is used to demonstrate exercises to the patient and to provide him feedback. 
We aim to develop a robot coach capable of understanding the requirements of a rehabilitation exercise from the medical expert's demonstrations. Then, it should be capable of demonstrating rehabilitation exercises to a patient. Currently, the learning of rehabilitation exercises is achieved by manually moving the robot in collaboration with physiotherapists and recording motor angles at each timestamps. Finally, the robot should watch him/her carry out the exercise and give him/her feedback so as to improve his/her performance and keep them motivated. To achieve these goals, human motion analysis is crucial. In that context we propose in this paper a multi-level human motion analysis to evaluate and assess rehabilitation movements performed in front of a RGB-D camera.

The rest of this paper is organized as follows: Section~\ref{relateWork} reviews existing approaches to address the problem of physical rehabilitation. Section~\ref{codesign} describes our methodology for the design of the system. Section~\ref{ITS} describes our proposed multi-level human motion analysis approach for both learning ideal rehabilitation movements and thoroughly assessing patients' movements. The approach is evaluated on several kinaesthetic exercises in Section~\ref{experiments}. Finally, Section~\ref{conclusion} concludes the paper and investigates possible future work.

\section{RELATED WORK}\label{relateWork}

\subsection{Coaching Robots for Physical Exercise}
Human motion analysis has been investigated in different contexts like action recognition, motion segmentation and fall detection. However, only few approaches addressed the challenge of physical rehabilitation through coaching robot systems. While several studies showed the potential of  virtual agents \cite{Waltemate2015PSVRSTV,Anderson2013ACE} and physical robots  \cite{Belpaeme2012JHI}  to enhance engagement and learning in health, physical activity or social contexts, Fasola \textit{et al} \cite{Fasola2013JHI} showed better assessment by the elderly subjects of the physical robot coach compared to virtual systems. Robots  for coaching physical exercises have been recently presented \cite{Gorer2013,Takenori2015}. However, Takenori \textit{et al} \cite{Takenori2015} provided no feedback or active guidance to the patient. 

%
%
\subsection{Representing the human body}
While in these approaches only joint angles are considered as motion features, other approaches also consider their Cartesian positions~\cite{Nguyen2016R2IISHRIC,Tao:2007} or relative transformations (translation and rotation)~\cite{wang2015} for more robustness to subjects' sizes and morphology variations. Relative position features are more suitable for exercise characteristics like ``Place your hands at head's height". Orientation features better correspond to characteristics like ``Stretch your arm horizontally". In our project we propose a combination of both relative position and orientation features. 
To represent human body data, Euclidean space is traditionally employed. However, an increasing number of approaches consider Riemann spaces to represent human postures and motions so as to handle the non linearity of human movements. Its proven effectiveness in human motion analysis for action recognition~\cite{vemulapalli2014,Devanne:2017} motivated us to adopt this framework.

\subsection{Modeling the ideal movement and acceptable variations}
In robotics, imitation learning 
has explored probabilistic methods based on Hidden Markov Models (HMM) and Gaussian Mixture Models (GMM) to enable robots to learn by observation of demonstrations such as in \cite{Calinon2007ITSC}. The GMM thus learned after demonstrations constitute a probabilistic description of the ideal movement with an adaptive acceptability measure of errors. The model is robust to noise and small errors in  the training data. In this work, we propose a similar approach based on GMM to learn a model representing ideal exercises and represented on a Riemannian space combining position and orientation features.

\subsection{Assessing movements}
Assessing the patient's performance to provide him adequate and personalized support to help him correct his errors is essential for an intelligent tutoring system. Takenori et al.\cite{Takenori2015} only focuses on learning a perfect movement and does not tackle the assessment of an imitation attempt. Goerer et al.~\cite{Gorer2017} based their automatic evaluation on the  distance measure between the user's current arm angles and the specified goal arm angles, based on only one template movement. The use of probabilistic models such as GMM are more suitable to analyze deviation according to an ideal movement and has been for instance successfully applied for abnormal gait detection~\cite{Devanne:2016}. In addition, 
spatial information about which body part is incorrect facilitates error understanding and improvement by the patient. In human motion analysis, movement segmentation is often adopted to face human motion complexity. 
In this work we propose to segment exercises online in motion primitives in order to locally analyze patients' movements. In addition, this allows to temporally localize errors which can be beneficial for the patient's understanding. Finally, automatically providing instructions on how to improve the movement can significantly help the patient perform the correct movement and keep him motivated throughout the rehabilitation session. We propose a multi-level analysis of human motion for assessing physical exercises in a context of a robot coach system for rehabilitation.
%
%

\section{Co-design with therapists and psychologists}\label{codesign}

In order to design our system and our HRI, we first choose an anthropomorphic robot platform, and led a psychological analysis of the target population.

\subsection{An anthropomorphic robot with a spinal cord}
In the literature on robot coach systems, the humanoid Nao robot is often used to achieve this task~\cite{schneider2016exercising,ramgoolam2014towards}. 
After first tests with Nao\footnote{http://keraal.enstb.org/media/videos/KeraalProofOfConcept.mp4}, we were quickly limited to only arm movements.  On the contrary, patients suffering from low-back pain need physical exercises to muscle the back and their lumbar spine. However, Nao has only one DOF for the trunk. This makes the performance of many rehabilitation exercises for low back pain impossible or not natural. 

Conversely, the Poppy robot is designed be anthropomorphic~\cite{Lapeyre2014} with 25 degrees of freedom (DOF) including a 5 DOFs articulated trunk. Given its unique capability of realizing movements of the lumbar spine, this robot fits well with the objectives of rehabilitation programs dedicated to low back pain. 
We also took advantage of the fact that Poppy is an open-source platform based on 3D printing to add wrists in order to carry out the exercises we identified with therapists.

However, Poppy only has a single 2D camera that does not allow easy motion analysis. This is why we complemented our system with a 
 Kinect camera, chosen for its low-cost and ease of use both for the therapists and the patients. It is all the more advantageous than it is a seamless sensor as no markers or specific suit are needed, nor time for setup.

\subsection{Acceptance by the target population}
Because rehabilitation patients are often elderly people, who are not used to new technologies, the issue about acceptability is even more acute than in other applications. This is why we carried out psychological tests with subjects from the target population: 5 subjects over 60 years old, including 2 women and 3 men, and including a rehabilitation patient. They use computers, tablets or online applications, but are not tech-savvy or scientists. We asked the subjects to perform 3 repetitions of 5 exercises after the demonstrations by Poppy Torso (we used only the upper body for this test). Afterwards, we interviewed them about their own focuses and perceptions and we found out three main shared concerns:

\begin{itemize1}
\item conformity issue : The patients were dedicated to executing the movements well and  aware of the difference between the robot's and the human movements. They were concerned about: does one's execution of the movement correspond to the desired movement? 
How can one interpret the demonstrations by the robot?
\item performance issue : the patients felt competing with the robot but also with themselves, feeling they had to prove something to themselves. How can the patient assess his performance? How this motivation be used for rehabilitation?
\item emotions issue : whereas they expected an intimidating robot, its physical aspect and motions were rated positively and friendly. How can a friendly robot entice motivation? Still, in the environment of the test the authority acknowledged by the patient to the robot was never questioned. They put themselves under pressure to copy the movements. How can demonstrations be accurate ?
\end{itemize1}

Results from this qualitative study (submitted) validate that our robotic platform can be used as a motivation tool for patients. They reveal the importance of precise movements, of clear instructions, but also the desire to have feedback on their performance. Thus we referred to therapists.

\subsection{Selection of the exercises}
We presented the robot and sensor system to therapists to determine whether our robot could coach the usual exercises. Based on the limitations of our robot and the Kinect sensor, we then selected seven exercises\footnote{Videos of exercises are available at: www.keraal.enstb.org/exercises.html}. Eventually for the clinical trials, we implemented only three exercises, that were using the same sitting position, so as to avoid position change. To make demonstrations clearer, we added oral instructions based on text-to-speech. We designed with the therapists the motions but also the instructions and the feedback for the HRI. In particular, to entice motivation for patients, the robot will give advice as to how to improve the movement, instead of only pointing out the errors. The system is currently tested in clinical trials for 4 months to include 30 rehabilitation patients.

\section{Multi-level Human Motion Analysis for Physical Rehabilitation}\label{ITS}
To guarantee an efficient intelligent tutoring system within our robot coach, two phases have been identified, the learning phase and the assessment phase. 

\subsection{Human Motion Learning  }
In robotics, 
Gaussian Mixture Models (GMM) have proven successful for robots learning by observation of demonstrations \cite{Calinon2007ITSC}. In our context of physical rehabilitation, expert demonstrations correspond to human motion sequences. Hence an efficient representation of human motion is needed. 


\subsubsection{Human Pose Space} 
\begin{figure}[!th]
\vspace{-0.2cm}
\centering
\subfloat[ ]{
	\includegraphics[width=0.45\columnwidth]{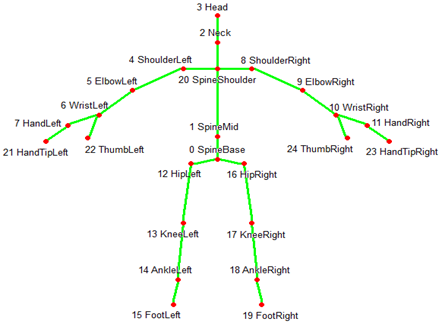}\label{skelFig}
    }
\subfloat[ ]{
	\includegraphics[width=0.3\columnwidth]{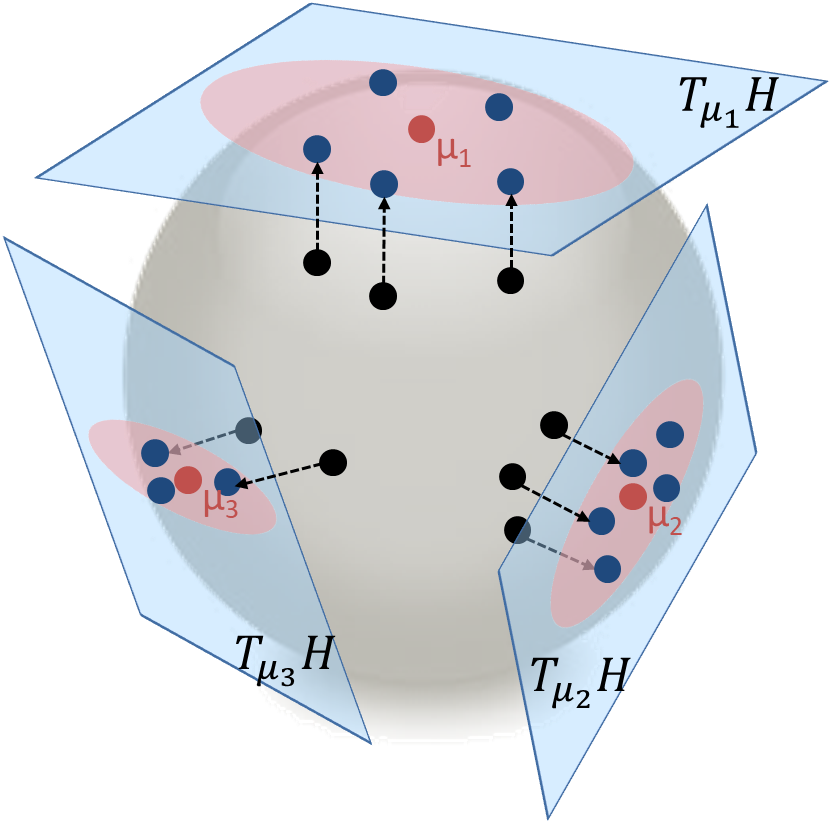}\label{GMMManifold}
    }
\vspace{-0.2cm}
\caption{\small Representation of the human pose.~\protect\subref{skelFig} The structure of the skeleton captured by Kinect.~\protect\subref{GMMManifold}Illustration of the human pose space $\mathcal{H}$ with three Gaussians computed on tangent space at means $\mu_k$ (red dots). Black dots are elements on the manifold and blue dots are their projection on tangent spaces.}\label{HumanPose}
\vspace{-0.4cm}
\end{figure}

The Microsoft Kinect provides in real-time the 3D position and the orientation of 25 joints forming a humanoid skeleton. Figure~\ref{skelFig} shows the structure of the skeleton detected using Kinect. In this work, we focus our analysis on upper body joints, using $J=11$ joints. 

To allow invariance to subjects' sizes and positions, we employ normalized relative positions. For a given joint $j$, its 3D cartesian position $P_j$ is computed relatively to the Spine Shoulder absolute position $p_{ss}$ and normalized using the length $L_{spine}$ of the spine bone (between Spine Shoulder joint and Spine Mid joint): $P_j = (p_j - p_{ss}) / L_{spine}$. As a result, a skeleton pose $y_t$ at frame $t$ can be represented as:
$$
y_t = [O_1, P_1, O_2, P_2, \dots, O_J, P_J ], \eqno{(1)}
$$
\noindent where $O_j$ is the orientation of joint $j$ and $P_j$ its  position.

 
While joint positions are naturally viewed in 3D Euclidean space, quaternions can be represented as elements of the 3-sphere $\mathcal{S}^3$ which is a 3 dimensional Riemannian manifold. A Riemannian manifold is a smooth space that locally resembles Euclidean space and is equipped with the Riemannian metric defined on the tangent space at each point of the manifold~\cite{jost2005riemannian}. Furthermore, the Cartesian product of several Riemannian manifolds is again a Riemannian manifold. This property allows us to consider combinations of joint quaternions and positions corresponding to the whole body. We define the human pose space $\mathcal{H}$ as the Cartesian product of quaternion and position of all skeleton joints:
$$
\mathcal{H} = \mathbb{R}^3 \times \mathcal{S}^3 \times \mathbb{R}^3 \times \mathcal{S}^3 \times \dots \times \mathbb{R}^3 \times \mathcal{S}^3 . \eqno{(2)}
$$

\subsubsection{Imitation Learning Algorithm}
For learning a model representing an ideal movement from several expert demonstrations, 
we have employed the recent framework proposed in~\cite{Zeestraten2017} extending common imitation learning techniques, such as GMM, to Riemannian manifolds. Such framework is particularly convenient for our work as our skeleton features are represented in the Riemannian human pose space.

A common way to handle the non-linearity of Riemannian manifolds $\mathcal{M}$ is to consider tangent spaces $\mathcal{T}_p\mathcal{M}$ at a reference point $p \in \mathcal{M}$ as a linear approximation of the neighborhood of $p$. To map a point $g$ from the manifold to the tangent space at $p$ resulting in $v$, the distance preserving logarithmic map is defined as Log$_p (.) : \mathcal{M} \rightarrow \mathcal{T}_p\mathcal{M}$. Conversely, the exponential map Exp$_p (.) : \mathcal{T}_p\mathcal{M} \rightarrow \mathcal{M}$ allows to go back from the tangent space to the manifold. More details about exponential and logarithmic map computation on the $\mathcal{S}^3$ manifold can be found in~\cite{Zeestraten2017}. As the human pose space $\mathcal{H}$ is the Cartesian product of several manifolds, corresponding exponential and logarithmic mapping are obtained by concatenating individual functions of each sub-manifold.

Using linear tangent spaces, we can compute approximated multivariate Gaussians on the human pose space. The mean $\mu$ of $N$ points $y$ on the human pose space can be obtained using~\cite{Karcher:1977}:
$$
\mu = \arg \min_p \sum_{i=1}^{N} d(p, y)^2, \eqno{(3)}
$$
\noindent where $d(p, y)$ is the geodesic distance on the manifold which can be written using logarithmic map as $d(p, y) = \|$Log$_{p}(y)\|$. Such mean is called the Riemannian center of mass~\cite{Karcher:1977} and is obtained by an iterative process until no change. 
Once a mean point is computed, the covariance matrix $\Sigma$ can be computed from points $y_i$ projected into the tangent space at $\mu$ using the logarithmic map Log$_{\mu}(y_i)$. We can then learn a Gaussian Mixture Model defined as:
$$
p(x) = \sum_{k=1}^{K} \phi_k \mathcal{N}(x | \mu_k, \Sigma_k), \eqno{(5)}
$$

\noindent where $x$ encodes both the human pose $y_t$ and the timestamps $t$, $K$ is the number of Gaussians, $\phi_k$ the weight of the $k$-th Gaussian, $\mu_k$ the Riemannian center of mass of the $k$-th Gaussian computed on the manifold and $\Sigma_k$ the covariance matrix of the $k$-th Gaussian. The parameters $\phi_k$, $\mu_k$ and $\Sigma_k$ are learned using Expectation-Maximization on the human pose Space~\cite{simo20163d}. Figure~\ref{GMMManifold} illustrates the human pose space $\mathcal{H}$ with three Gaussians computed on tangent space at means. We note that expert demonstration are first temporally aligned using dynamic programming. This allows us to handle possible velocity variations among expert demonstrations.



\subsubsection{Ideal movement generation}
Once a model is learned for each exercise, we can generate an optimal sequence using Gaussian Mixture Regression (GMR) which approximates the sequence using a single Gaussian:
$$
p(\hat{x}|t) \approx \mathcal{N}(\hat{\mu}, \hat{\Sigma}) . \eqno{(6)}
$$
As described in~\cite{Zeestraten2017}, to remedy the non linearity of the manifold, $\hat{\mu}$ is computed in an iterative process similarly to the Riemannian center of mass and $\hat{\Sigma}$ is computed on the tangent space at $\hat{\mu}$. By evaluating $\hat{x}$ for successive values of $t$, we obtain a generalized form of the ideal motion $\hat{X}$. This optimal motion sequence will be used as a reference to evaluate  a motion sequence of a patient.

\subsection{Human motion assessment}
\begin{figure}[!th]
\vspace{-0.4cm}
\centering
\includegraphics[width=0.6\columnwidth]{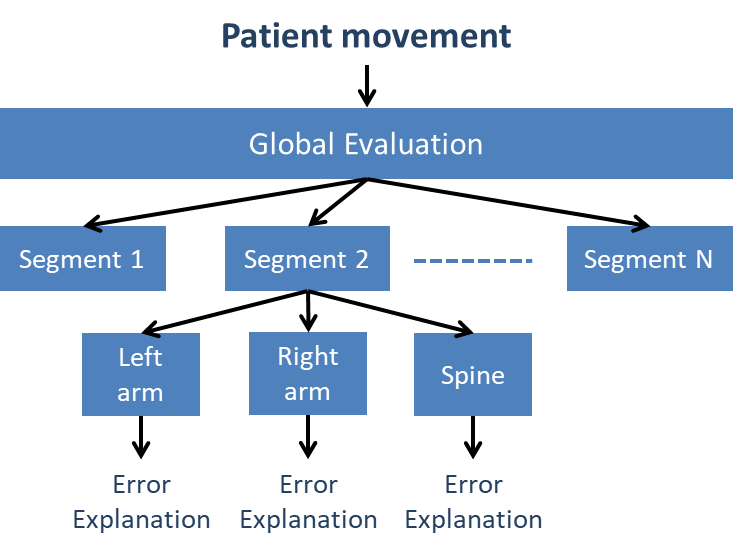}
\caption{\small Illustration of the multi-level analysis of patients' movement. }\label{multiLevelAnalysis}
\vspace{-0.6cm}
\end{figure}

In order to evaluate a test sequence of a patient and provide him feedback, we propose a multi-level analysis of the movement, as illustrated in Figure~\ref{multiLevelAnalysis}.

\subsubsection{Global evaluation}
To evaluate a patients' movement we first temporally aligned the motion sequence to the ideal movement computed above using dynamic programming. Then, we compute the log-likelihood that the given sequence $X$ has been generated by the learned Gaussian Mixture Model of the corresponding exercise:
$$
ln(p(X | \phi, \mu, \sigma) = \sum_{t=1}^T ln(\sum_{k=1}^K \phi_k \mathcal{N}(x_t | \mu_k, \Sigma_k)) . \eqno{(7)}
$$
As the log-likelihood value may not be significant for a patient or for a physiotherapist, we can use thresholds to translate the log-likelihood into percentage of success.

\subsubsection{Temporal segment analysis}
\begin{figure}[!th]
\vspace{-0.4cm}
\centering
\includegraphics[width=0.48\columnwidth]{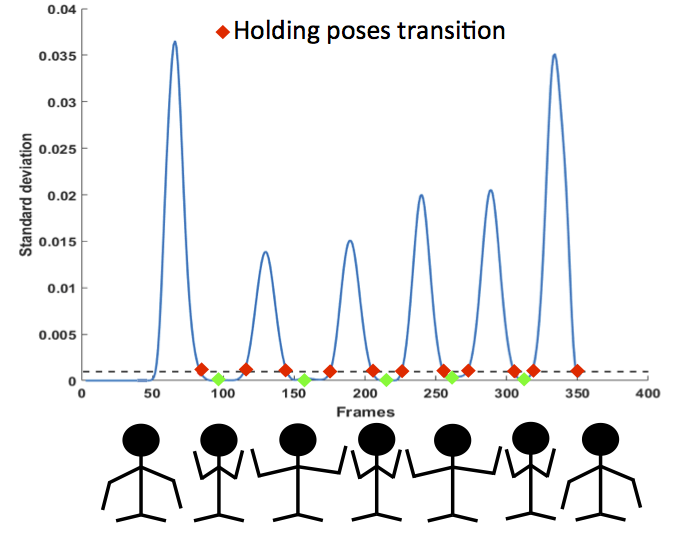}
\caption{\small Motion evolution for segmentation
 by differentiating transition motions and holding postures. Red points correspond to boundaries between two temporal segments. Green points correspond to the beginning of the exercise. The black line corresponds to the threshold used to detect starting frames and boundaries in the second strategy.}\label{segmentation}
\vspace{-0.6cm}
\end{figure}

While providing a global score for the sequence allows the patient to have a global idea on his performance, it is also interesting to know which part of the exercise is not performed correctly. Thus, we propose an analysis based on temporal segments. 
We analyze the motion within a window of length $W$ by computing the standard deviation among data included in the temporal window: 
$$
\sigma =  \sqrt{\frac{1}{W}\sum_{t=1}^{W} d(\mu_W, y_t)^2}, \eqno{(8)}
$$
\noindent where $\mu_W$ is the Riemannian center of mass of all skeleton poses of the sequence included in the window. 
We can detect key frames when the motion value crosses the threshold. With this strategy, we are able to differentiate transition movements and holding postures. According to the evaluated exercise, we can select the corresponding segmentation strategy. Figure~\ref{segmentation} shows an example of exercise segmentation using our two strategies. Detected key frames representing boundaries between temporal segments are depicted in red color.

In addition, by detecting when motion value crosses the threshold for the first time, we can identify the beginning of the exercise. We select the beginning $10$ frames before such crossing point. This allows us to compare sequences starting at the same time. The detected starting frames are depicted in green color in Figure~\ref{segmentation}.

\subsubsection{Body part analysis}
We also propose a local analysis to differentiate different body parts. This allows to identify which body part is more responsible of the error and gives more precise feedback to the patient. As the skeleton pose $x_t$ is the concatenation of all joint features, we can compute the log-likelihood for data corresponding to joints of the desired body part to obtain a score for each body part. As shown in Figure~\ref{multiLevelAnalysis}, we differentiate the two arms and the spine. 

\subsubsection{Improvement advice}
Until now, the evaluation of a motion sequence only provides a global score of success. However in the case a movement is not correctly performed, it is very important to understand the reason why it has been detected as incorrect and propose a solution how to improve the movement. When an error is detected for a given body part, we propose to classify this error in order to infer the corresponding advice. 
For each pose, we made a projection of the skeleton pose $x_t$ on the tangent space of $\mathcal{H}$ of the ideal movement $\hat{X}$. The projection $y_t$ thus represents the distance of the measured pose $x_t$ to the ideal movement's pose $\hat{x}_t$. We then classify this error $y_t$, using a SVM, to the errors we have already identified. If the confidence is high, the robot uses a dictionary to orally give the preset advice.

\section{Experimental Results}\label{experiments}
We propose to evaluate our method on the three rehabilitation exercises. In the first exercise, the patient raises his arms horizontally and then rotates the trunk on both sides. In the second exercise, the patient should raise an arm upright then lean on the side. The same is the performed for the other arm. In the third exercise, the patient should lift his arms in front of him with the elbows bent and then spread the arms.
%
%


\subsection{Dataset}
Under the supervision of a physiotherapist, we collect a training database of two different subjects performing each exercise three times. Then we collect test data of a  third subject performing the same exercises twice. In addition, this test subject performs incorrect exercises by simulating errors. For the first exercise, the arms are not enough raised. For the second exercise, the subject does not tilt the arm and keep it straight. In the third exercise, the arms are not enough raised. Therefore, for each exercise, we have both test data which are correct and incorrect~\footnote{Videos available at: www.keraal.enstb.org/incorrectexercises.html}.

\subsection{Global Assessment}
First, we compute the global and body part scores for each test sequence. Results are reported in Table~\ref{scoreTable}. We can first observe that the obtained scores for correct exercises are much higher than those obtained for incorrect exercise. This shows that our method is able to detect when an exercise is not correctly performed. Moreover, as different subjects are used for training and testing, it shows that our method is independent of subjects. In addition, we can see that correct exercise 1 obtained lower scores than other correct exercises. This can be explained by the fact that when the subject is rotating the trunk, one arm is behind the chest and may be incorrectly detected by the Kinect sensor. Thus it can affect the overall score but not as much as incorrect exercises. Finally, by observing the scores of the spine, we can see that a score of 100\% is obtained for most cases. This is expected as the spine should always be straight. However, for the incorrect exercise 2, when the subject keeps his back straight and does not lean on the side, we can see that the score for the spine is 90\%. This shows that we are able to detect the error. However, in comparison with other errors in the arms, the score is still high. This shows that errors in different body parts do not affect the score similarly. Adding some weights for each body parts could allow us to overcome this limitation. This will be part of our future work.

\begin{table}[h]
\caption{Evaluation results for all test sequences. Global scores and body part scores are reported for each sample.}
\label{scoreTable}
\vspace{-0.2cm}
\begin{center}

\begin{tabular}{|l|c|c|c|c|}
\hline
 & Global & Left arm & Spine & Right arm \\
\hline
Ex1 Correct & 72\% & 75\% & 100\% & 63\% \\
\hline
Ex1 Correct & 82\% & 86\% & 100\% & 73\% \\
\hline
Ex1 Error & \bf12\% & \bf31\% & 100\% & \bf19\% \\
\hline
Ex2 Correct & 92\% & 98\% & 100\% & 96\% \\
\hline
Ex2 Correct & 98\% & 100\% & 100\% & 99\% \\
\hline
Ex2 Error & \bf24\% & \bf37\% & \bf90\% & \bf49\% \\
\hline
Ex3 Correct & 95\% & 90\% & 100\% & 100\% \\
\hline
Ex3 Correct & 93\% & 99\% & 100\% & 90\% \\
\hline
Ex3 Error & \bf27\% & \bf47\% & 100\% & \bf40\% \\
\hline
\end{tabular}
\end{center}
\vspace{-0.2cm}
\end{table}

\subsection{Assessing each temporal segment}
\begin{figure}[!th]
\vspace{-0.2cm}
\centering
\includegraphics[width=0.9\columnwidth]{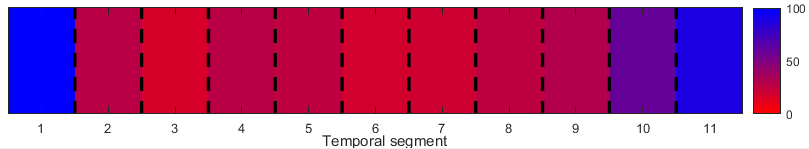}
\caption{\small Temporal evaluation. Scores are computed for each temporal segment. Red colors mean low scores, while blue colors represent correct scores. Black dot lines represents boundaries between temporal segments.}\label{scoreSegment}
\vspace{-0.4cm}
\end{figure}

In a second time, we compute scores for each temporal segment according to the two segmentation strategies. Figure~\ref{scoreSegment} illustrates results for the incorrect exercise 3. By using the first segmentation strategy where only transition motions are considered, we can see that all temporal segments are detected as incorrect but the first one and last one obtained higher scores. For the first segment, this is because the movement of raising arms is correct most of the time except near the end as the arms are not enough raised. As a result the mean score of this temporal segment is quite low. The same applies for the last segment where the subject lowers his arms. Only the beginning of the segments is incorrect as it starts with the arms too low. In comparison, if we use the second segmentation strategy, the first and last segments are not detected as incorrect. This is because the second strategy differentiates the movement and the holding postures. This shows that for this third exercise, the second strategy is more suitable. It allows to detect the error on the holding postures and not on the previous transition motion which is correct.
%
%
%
%
%
%

%
%

\begin{figure}[!th]
\vspace{-0.4cm}
\centering
\includegraphics[width=0.49\columnwidth]{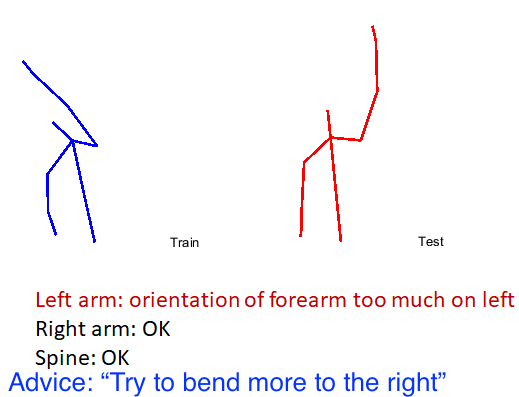}
\includegraphics[width=0.49\columnwidth]{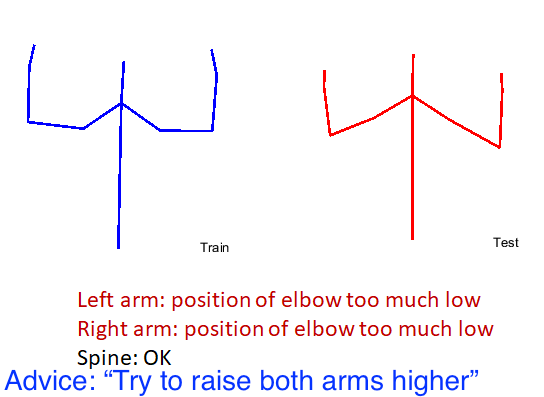}
\vspace{-0.4cm}
\caption{\small Error explanation \& oral advice (ex. 2:left, 3: right).}\label{explanationError}
\vspace{-0.4cm}
\end{figure}

Finally, we show two examples of explanation of detected errors for exercises 2 and 3. These examples are illustrated in Figure~\ref{explanationError}. Explanation sentences are build automatically according to the evaluation of the patient's movements. In our rehabilitation scenario, such sentences are sent to Poppy which enunciates them using a Text-To-Speech system. 

These experimental results show that our method is able to assess patients' exercises and provide corresponding feedback indicating where and why the movement is not correct.

\section{CONCLUSIONS}\label{conclusion}
We  have proposed a human motion analysis method for physical exercises assessment using motion segmentation to provide temporal analysis.
Our method learns a probabilistic model of position and orientation features from expert demonstrations. It considers variations among demonstrations and identifies which part of the exercise is important. In addition, our multi-level analysis allows to provide detailed feedback of detected errors including the body parts, the temporal segment and how to correct the error. Evaluation on three different rehabilitation exercises targeting low back pain demonstrated promising results.

The system is the result of the work we have undertaken with therapists and psychologists in order to validate our robotic platform, identify physical rehabilitation exercises and to design our HRI. The system is since November in clinical tests with low back pain patients within their rehabilitation program. These experiments will allow us to gather real-user data and evaluate our method with a larger set of users. Moreover, the acceptability of the Poppy robot as a robot coach will also be analyzed.

\addtolength{\textheight}{-12cm}   






{\footnotesize

\bibliographystyle{./IEEEtran} 
\bibliography{./Keraal}
}

\end{document}